**Impact of Corrosion on the Emissivity of Advanced Reactor Structural Alloys**

J. L. King[1,2,3], H. Jo[2*], A. Shahsafi[1*]

K. Blomstrand[2], K. Sridharan[2]

M. A. Kats[1†]

[1]University of Wisconsin-Madison, Department of Electrical and Computer Engineering

[2]University of Wisconsin-Madison, Department of Engineering Physics

[3]Jensen Hughes, Power Services Group

*These authors contributed equally to this work. †Corresponding Author: mkats@wisc.edu

# 1 Abstract

Under standard operating conditions, the emissivity of structural alloys used for various components of nuclear reactors may evolve, affecting the heat transfer of the systems. In this study, mid-infrared emissivities of several reactor structural alloys were measured before and after exposure to environments relevant to next-generation reactors. We evaluated nickel-based alloys Haynes 230 and Inconel 617 exposed to helium gas at 1000 °C, nickel-based Hastelloy N and iron-based 316 stainless steel exposed to molten salts at 750-850 °C, 316 stainless steel exposed to liquid sodium at 650 °C, and 316 stainless steel and Haynes 230 exposed to supercritical $CO_2$ at 650 °C. Emissivity was measured via emissive and reflective techniques using a Fourier transform infrared (FTIR) spectrometer. Large increases in emissivity are observed for alloys exposed to oxidizing environments, while only minor differences were observed in other exposure conditions.

# 2 Introduction

To achieve higher efficiencies, next-generation reactor concepts are targeting operating temperatures significantly higher than those of existing light-water reactors (LWRs) [1]. At these high operating temperatures, thermal radiation becomes a significant heat transfer mechanism, because radiated power from a hot surface increases as the fourth power of its temperature, according to the Stefan-Boltzmann relationship [2]. Presently, heat transfer models for high-temperature reactors such as molten salt reactors (MSRs) and high temperature gas-cooled reactor (HTGRs) use rough approximations for the emissivity of surfaces, which may be insufficient for accurate estimations of reactor component temperatures.

For example, preliminary calculations of worst-case shutdown afterheat scenarios (i.e., loss of salt from both the primary and secondary loops) for the molten salt breeder reactor (MSBR) yielded temperature differences of nearly 200 °C for differences in heat-exchanger emissivity of 0.1 [3]. In 2007, expert panels classified reactor pressure vessel (RPV) and reactor cavity cooling system (RCCS) emissivities as highly important for modular HTGRs but not adequately studied [4]. In HTGR systems and the associated RPVs and RCCSs, as much as 80% of overall RPV-to-RCCS heat transfer may be radiative during transient scenarios [5].

Several experimental studies of emissivity of various nuclear reactor alloy surfaces have recently been conducted, with exposure performed under controlled ex-situ conditions. In these studies, roughened surfaces were produced using abrasive blasting [6–8] and grinding papers [9–11], while surface oxidation was produced in high-temperature air [8–14]. Graphite coatings were manually deposited to simulate films expected to form in graphite-moderated gas reactors [6,7,9–11]. However, no studies have been performed on nuclear reactor alloys exposed to their intended high-temperature environments, with the exception of low-alloy steels in high-temperature air [13,15]. The purpose of this study is to fill the gap between surrogate environment exposures and exposure in the intended reactor environment.

## 2.1 Equations Governing Radiative Heat Transfer

Radiated power from a hot surface increases as the fourth power of its temperature, according to the Stefan-Boltzmann relationship [2]:

$$P = \varepsilon(T) A \sigma T^4 \quad (1)$$

where $P$ is radiated power [W], $\sigma$ is the Stefan-Boltzmann constant of $5.67 \times 10^{-8}$ [W·m$^{-2}$·K$^{-4}$], $A$ is the area [m$^2$] and $T$ is temperature [K].

The amount of heat radiated from a surface at a particular temperature is determined by its total hemispherical emissivity, $\varepsilon(T)$, defined as the ratio of the actual emitted power to that of a black body with the same surface area at the same temperature. The spectral and temperature-dependent nature of the emitted radiation is represented through the blackbody spectral radiative intensity, $I_{BB}(\lambda, T)$, and the spectral emissivity of the object, $\varepsilon(\lambda, T)$ where $\lambda$ is free-space wavelength [m]. The blackbody spectral radiative intensity is given by the Planck distribution [16]:

$$I_{BB}(\lambda, T) = \frac{2hc^2}{\lambda^5 \left(e^{\frac{hc}{k\lambda T}} - 1\right)} \ [\text{W·sr}^{-1} \cdot \text{m}^{-3}] \quad (2)$$

where $h$ is Planck's constant $\sim 6.63 \times 10^{-34}$ [J·sec], $c$ is the speed of light $\sim 3 \times 10^8$ [m·s$^{-1}$], and $k$ is Boltzmann's constant $\sim 1.38 \times 10^{-23}$ [J·K$^{-1}$]. The spectral directional emissivity, $\varepsilon(\lambda, \theta, \varphi, T)$, is defined by

$$\varepsilon(\lambda, \theta, \varphi, T) = \frac{I(\lambda, \theta, \varphi, T)}{I_{BB}(\lambda, T)}, \quad (3)$$

where $I(\lambda, \theta, \varphi, T)$ is the directional spectral radiative intensity of the object, and $I_{BB}(\lambda, T)$ is independent of direction. The spectral emissivity is found by integrating the spectral directional emissivity over all directions:

$$\varepsilon(\lambda, T) = \frac{1}{\pi} \int_{\varphi=0}^{2\pi} \int_{\theta=0}^{\pi/2} \varepsilon(\lambda, \theta, \varphi, T) \cos(\theta)\sin(\theta) \, d\theta d\varphi, \qquad (4)$$

Together, Eqns. 2 and 4 may be used to obtain the total hemispherical emissivity:

$$\varepsilon(T) = \frac{\int_{\lambda=0}^{\infty} \varepsilon(\lambda,T) I_{BB}(\lambda,T) d\lambda}{\int_{\lambda=0}^{\infty} I_{BB}(\lambda,T) d\lambda} \qquad (5)$$

$\varepsilon(T)$ may then be used in the Stefan-Boltzmann relationship (Eq. 1).

For simplicity, in this work we measure emission into air rather than into some working fluid, before and after exposure to reactor conditions. These measurements allow us to determine whether the emissivity is expected to change during reactor operation. Furthermore, measurements of emissivity into air can be directly used in scenarios where the medium is drained and replaced by air or vacuum, or in the case that the medium has optical properties similar to that of air. Though in-situ test reactor exposures can be prohibitively expensive or impractical, scaled-down facilities such as flow loops or crucibles provide suitable surrogates for exposure to media such as helium gas [17,18], molten salts [19–21], liquid sodium [22], and supercritical $CO_2$ ($sCO_2$) [23]. We focused on these four relevant environments, and one or two candidate alloys that are being considered for each environment. The temperatures of these environments were selected based on proposed steady state operating conditions for the respective reactor types. All samples were exposed in a flowing medium except for the molten-salt experiments. The sample information and exposure parameters are provided in Table 1. For the molten salt environments, hydrofluorination was used to purify the salt [24]. Note that impurity measurements are more difficult for the molten salt than for the other environments studied, and we are unable to report these impurity levels.

Table 1. Summary of test materials and exposure environments and conditions

| Reactor Application | Test material | Environment | Temperature, duration |
|---|---|---|---|
| Reference samples | Haynes 230 | - | - |
| | Inconel 617 | - | - |
| | Hastelloy N | - | - |
| | 316 stainless steel | - | - |
| HTGR core | Haynes 230 | He gas, carburizing | 1000 °C, 500 hr |

| | | <0.5 ppm H₂O, 10 ppm CO, 100 ppm CH₄ | |
| --- | --- | --- | --- |
| structure & cooling systems | | He gas, oxidizing 4 ppm H₂O, 40 ppm CO, 20 ppm CH₄ | 1000 °C, 500 hr |
| | Inconel 617 | He gas, carburizing <0.5 ppm H₂O, 10 ppm CO, 100 ppm CH₄ | 1000 °C, 500 hr |
| | | He gas, oxidizing 4 ppm H₂O, 40 ppm CO, 20 ppm CH₄ | 1000 °C, 500 hr |
| Molten salt reactor (MSR) core structure & cooling systems | Hastelloy N | Molten LiF-KF-NaF No flow | 850 °C, 1000 hr |
| | 316 stainless steel | Molten LiF-BeF₂ No flow | 750 °C, 1000 hr |
| Sodium fast reactor (SFR) cladding, structure & cooling systems | 316 stainless steel | Liquid sodium 10 ppm O₂ Flow rate ~ 6 m/s | 650 °C, 1000 hr |
| Advanced reactor sCO₂ heat exchanger | Haynes 230 | Research grade CO₂ H₂O < 3ppm, N₂ < 5 ppm, THC < 1 ppm, Ar + O₂ + CO < 1ppm Flow rate ~ 0.11 kg/hr 20 MPa | 650 °C, 400 hr |
| | 316 stainless steel | Research grade CO₂ H₂O < 3 ppm, N2 < 5 ppm, THC < 1 ppm, Ar + O₂ + CO < 1ppm Flow rate ~ 0.11 kg/hr 20 MPa | 650 °C, 400 hr |

Four alloys were tested in total, with their nominal compositions listed in Tables 2 and 3. Bal. refers to the remaining balance of composition after accounting for the other elements.

Table 2. Nominal compositions (%) of the Ni-based alloys tested

| Alloy | Fe | Cr | Ni | Co | Mo | Al | Mn | Si | C | W | Ti |
| --- | --- | --- | --- | --- | --- | --- | --- | --- | --- | --- | --- |
| Hastelloy N | <5 | 7 | Bal. | <0.2 | 16 | <0.35 | <0.8 | <1 | <.08 | 0.5 | <0.35 |
| Haynes 230 | <3 | 22 | Bal. | <5 | 2 | 0.3 | 0.5 | 0.4 | 0.1 | 14 | <0.1 |
| Inconel 617 | 1.6 | 22.2 | Bal. | 11.6 | 8.6 | 1.1 | 0.1 | 0.1 | 0.05 | - | 0.4 |

Table 3. Nominal compositions of the Fe-based alloy tested

| Alloy | Fe | Cr | Ni | Cu | N | P | S | Mo | Mn | Si | C |
|---|---|---|---|---|---|---|---|---|---|---|---|
| 316 stainless steel | Bal. | 16.83 | 10.03 | 0.38 | 0.05 | 0.03 | <0.01 | 2.01 | 1.53 | 0.31 | .0225 |

# 3 Emissivity Measurement Techniques

Two techniques were used in this study to measure emissivity. The first involved directly measuring the thermally emitted intensity from the samples, which were maintained at temperature $T = 300$ °C, and comparing it to that of a reference. The second involved measuring specular reflectance, and then using Kirchhoff's law of thermal radiation [25] to calculate the emissivity. These techniques provide equivalent results for samples that are opaque, optically flat (*i.e.*, non-scattering), and in equilibrium (*i.e.*, the measured region is at a single, well-defined temperature). For samples that may not satisfy some of these conditions, the reflection method can yield an incorrect emissivity. We used the direct-emission technique for all samples, except for the LiF-BeF$_2$-exposed sample, which was too small for our direct-emission apparatus.

The spectral range of our direct emission technique was limited to wavelengths of $3 < \lambda < 18$ μm, while the reflectance technique was limited to $1.25 < \lambda < 17$ μm. Since thermal emission occurs across all wavelengths, it is useful to consider the significance of each wavelength range across a range of relevant temperatures. Fig. 1 provides the ratio of radiant power within the measured ranges, $P_{MR}$, to the total radiant power, $P_T$, across all wavelengths for a blackbody, as a function of temperature. Both of our experimental spectral ranges contain more than half of the blackbody radiant power at temperatures from 0 °C to 1000 °C.

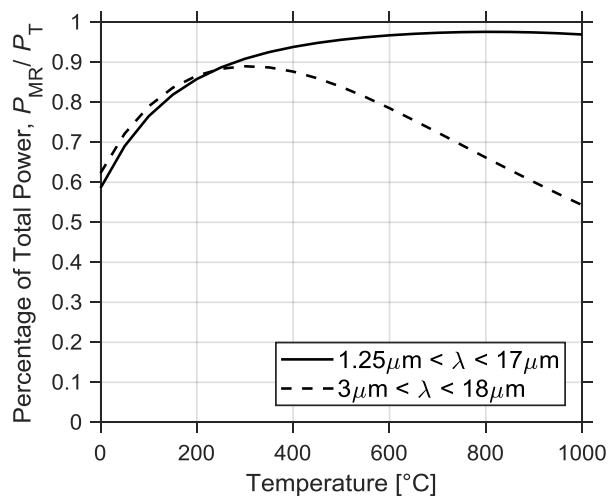

Fig. 1. The percentage of total blackbody radiant power within our experimental spectral ranges.

## 3.1 Direct-Emission Measurement Setup

For the direct-emission measurement, we used a Bruker Vertex 70 Fourier transform infrared (FTIR) spectrometer with a liquid-nitrogen-cooled mercury-cadmium-telluride (MCT) detector. We assume the sample emission signal $S$ measured by our FTIR can be decomposed as:

$$S(\lambda, T) \approx \eta(\lambda)\varepsilon(\lambda, T)I_{BB}(\lambda, T) \quad (6)$$

$$\eta = \text{System response [arb. units]}$$

Eqn. 6 is an approximation that ignores the contribution of thermal emission from the cooled detector and the room-temperature components of the instrument, which can sometimes be considerable [26]. We found that these contributions do not significantly affect our measurement for sample temperatures of 300 °C (see supplementary section 1.1.1 [27]), so this temperature was used for all measurements. By comparing the thermal emission from our samples to that of reference at the same temperature, $S_R(\lambda, T)$, which has a known emissivity $\varepsilon_R(\lambda, T)$, we extracted the sample emissivity as:

$$\varepsilon(\lambda, T) = \varepsilon_R(\lambda, T)\frac{S(\lambda, T)}{S_R(\lambda, T)} \quad (7)$$

Our reference was a silicon wafer with a coating of vertically-aligned carbon nanotubes (CNTs, height = 100 μm), with a measured emissivity, $\varepsilon_R(\lambda, T)$, of .90 +/- .02 that was assumed to not change significantly with wavelength. All measurements were performed at near-normal incidence, and angular dependence is not considered. A detailed analysis of error and uncertainty in our methodology, including the verification of the reference emissivity, is provided in section 1.1 of our supplementary [27].

During the emission measurement, samples were vertically affixed to a heated stage using Kapton tape, and a highly conductive thermal grease (Timtronics Red Ice 611HTC) was used to ensure good thermal contact with the stage. Measurements were taken at near-normal incidence with an approximate numerical aperture (NA) of 0.003.

## 3.2 Reflectance Measurement

One molten salt-exposed sample was too small to be measured by the direct-emission setup, so measurement via reflectance was pursued. We used Kirchhoff's law for opaque, non-scattering objects to convert reflectance to emissivity:

$$\varepsilon(\lambda, \theta, \varphi, T) = 1 - R(\lambda, \theta, \varphi, T) \quad (8)$$

Our reflectance measurement apparatus had a spot size small enough (about 100 μm in diameter) to measure the salt-exposed sample. However, the spot was also so small that any localized surface heterogeneities could skew the results if only one measurement was taken. Therefore, four measurements were taken at different positions on the sample, and an average of the measurements is presented. The Bruker Vertex 70 FTIR was once again used, this time coupled to a Hyperion microscope, using a reflective objective with a numerical aperture of 0.4. Analysis of errors and uncertainty for this approach is provided in section 1.2 of the supplementary [27].

# 4 Sample Preparation and Results

There are several mechanisms by which exposure to various high-temperature environments can modify the emissivity of alloy surfaces. A change in surface chemical composition from exposure can result in a large change in the optical constants of the surface, thereby changing the emissivity. If the chemical change in the surface results in a formation of one or more thin layers, thin-film interference can significantly affect the spectrum [28,29]. For films of sufficient thickness and optical absorption, the emissivity will simply be the bulk emissivity of the top layer. Exposure may also result in rougher surfaces, which can modify the emissivity [30,31]. We used a ZYGO white-light interferometer to measure the RMS roughness of each sample; note that initially, each sample was ground to a mirror finish using silicon carbide (SiC) papers.

Within the measured emissivity spectra, features due to $CO_2$ absorption ($\lambda \sim 4.2$ μm) [32] were observed in the spectra of most samples. Fortunately, this region is isolated to a narrow range of 4.18 μm $< \lambda <$ 4.4 μm. We assumed that the intrinsic emissivity spectra did not have sharp features in this range and thus we excluded the data points in this range, and connected the remaining data with a straight line from the data point at $\lambda = 4.18$ μm to the point at $\lambda = 4.4$ μm. Sharp features in the 5-8 μm range are due to atmospheric absorption by $H_2O$ [32].

## 4.1 Helium Gas Environment

Nickel-based alloys Inconel 617 and Haynes 230 are the leading candidate metallic alloys for the construction of HTGRs, where it is likely that helium will be used in the primary loop of the reactor, and outlet temperatures may reach as high as 950 °C during steady state operation [33]. In this work, samples were exposed in Idaho National Laboratory's high-temperature helium gas flow loop at 1000 °C for 500 hours [17], with exposure conditions designed to result in either oxidation or carburization of the alloys. The oxidative environment contained helium with impurity levels of 4 ppm $H_2O$, 40 ppm CO and 20 ppm $CH_4$, while the carburizing environment contained helium with < 0.5 ppm $H_2O$, 10 ppm CO and 100 ppm $CH_4$.

The addition of $H_2O$ to the helium gas yielded a strongly oxidative environment for both nickel-based alloys, Inconel 617 and Haynes 230. After exposure, both materials formed thick, black oxide layers (the presence of the oxide was verified using backscattered electron micrograph analysis, conducted on samples exposed to similar conditions at the same facility [34]), which resulted in a large increase in emissivity (Fig. 2). For both oxidizing-helium-exposed nickel-based alloys, as well as sCO2-exposed Haynes 230 presented in section 5.4, strong spectral features appear between $16 < \lambda < 18$ μm. The surface oxide layer of these samples is expected to be predominantly $Cr_2O_3$ [34], which exhibits two high-intensity absorption bands between 15 and 18 μm [32]. Our energy-dispersive x-ray spectroscopy (EDS) measurements, provided in Section 3 of the supplementary [27], indicate the presence of a chromium oxide on the surfaces of these samples.

In the carburizing environment, reactions of Cr with CO or CH$_4$ are expected to produce chromium carbide (assumed to be Cr$_{23}$C$_6$) [35]. Chromium carbide is also indicated in our supplementary EDS measurements [27]. The carburizing environment yielded large increases in roughness (see Fig. 2 and Fig. 3 captions). Overall, a moderate increase in emissivity is observed for both the Haynes 230 (Fig. 2) and the Inconel 617 (Fig. 3) alloys.

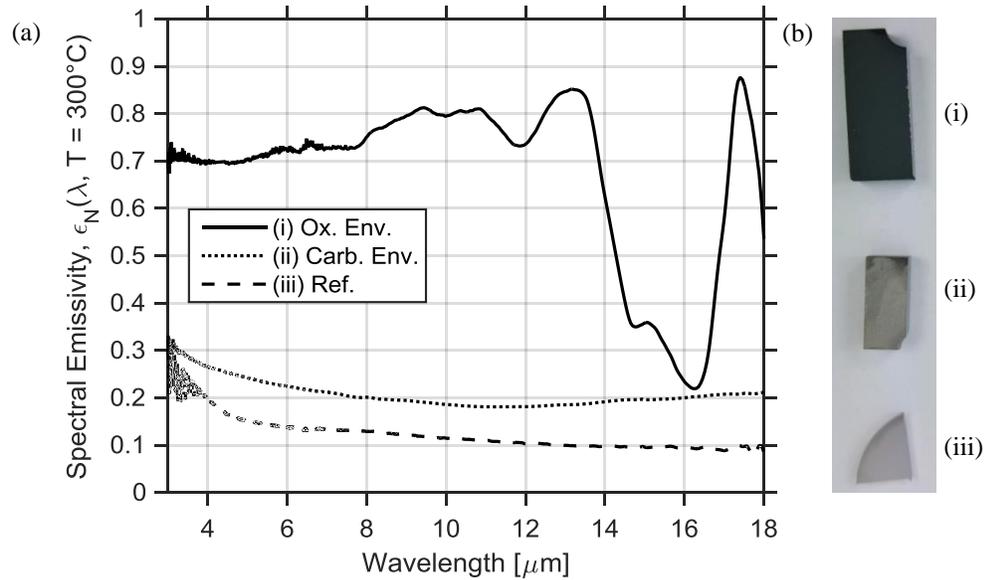

Fig. 2. (a) Spectral normal emissivity of Haynes 230 before (dashed) and after 500-hour exposure to 1000 °C helium gas with oxidizing (solid) and carburizing (dotted) chemistry. The measurements were taken at 300 °C. (b) Photos of the samples. The RMS roughness of the oxidizing-helium-exposed sample was 0.9 µm, compared to 1.86 µm for the carburizing-helium-exposed sample, and 0.03 µm for the unexposed reference.

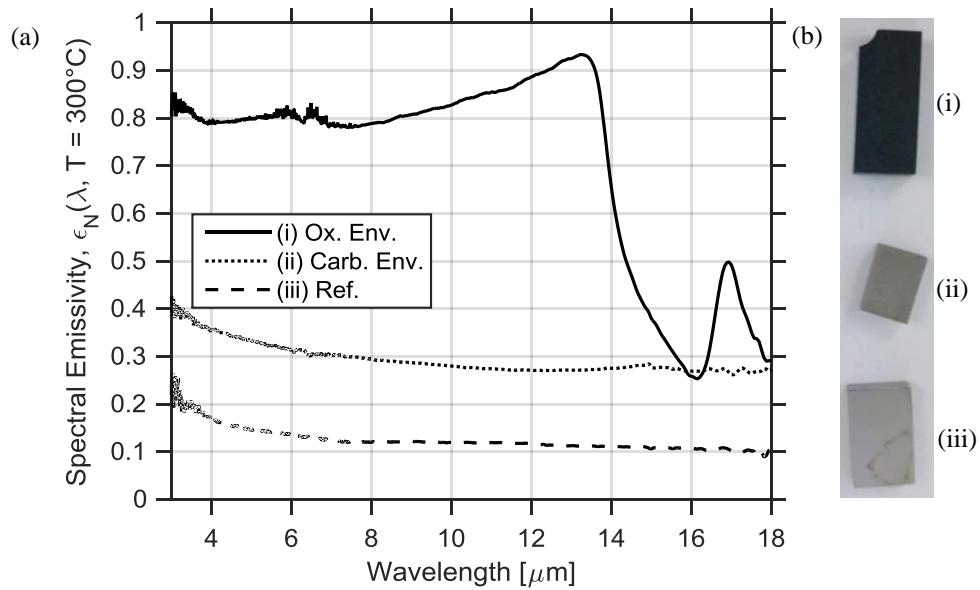

Fig. 3. (a) Spectral normal emissivity of Inconel 617 before (dashed) and after 500-hour exposure to 1000 °C helium gas with oxidizing (solid) and carburizing (dotted) chemistry. Measurements taken at 300 °C. (b) Photos of the samples. The RMS roughness of the oxidizing-helium-exposed sample was 1.6 µm, compared to 1.7 µm for the carburizing-helium-exposed sample, and 0.02 µm for the unexposed reference.

## 4.2 Molten Salt Environment

Molten salt reactors are expected to operate at temperatures of 700 °C or higher [1]. We studied Hastelloy N exposed to high-purity molten LiF-NaF-KF (FLiNaK) at 850 °C for 1000 hours in a 316 stainless steel crucible, which was procured from an earlier study at the University of Wisconsin-Madison [19]. Additionally, we also investigated a 316 stainless steel sample exposed to molten $LiF-BeF_2$ (FLiBe) for 1000 hours at 700 °C in a 316 stainless steel crucible from a separate earlier study [18]. 316 stainless steel has emerged as a leading candidate for the construction of the vessel in fluoride salt-cooled high-temperature reactors (FHRs). FLiBe is being considered as the primary coolant for the FHR for non-fueled coolants, and also forms the basis for fuel-dissolved salts [36].

The most significant corrosion mechanism in structural alloys in molten fluoride salts is chromium depletion [19]; this process is exacerbated because the oxide layer that helps protect the surface in other high-temperature environments is generally unstable in molten fluoride salts [36]. Therefore, Hastelloy N, which was specifically developed for molten fluoride salt applications [37,38], contains only about 6-7% Cr. Upon exposure to FLiNaK, Hastelloy N experiences relatively little corrosion, and the emissivity remains mostly unchanged (Fig. 4).

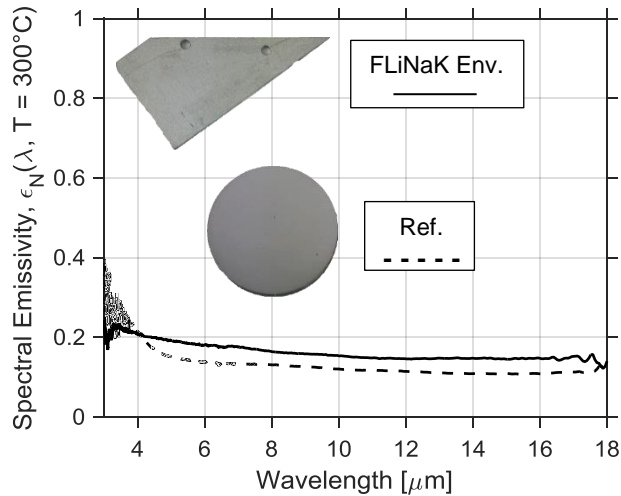

Fig. 4. Spectral normal emissivity for Hastelloy N before and after exposure to 850 °C FLiNaK for 1000 hours. Measurements taken at 300 °C. Photos of the samples are provided with the background cropped out. The RMS roughness of the FLiNaK-exposed Hastelloy N was 1.5 µm, compared to 0.03 µm prior to exposure.

316 stainless steel contains greater concentration of chromium (~17%) than Hastelloy N, so chromium depletion for this material occurs more readily. An overview of 316 stainless steel corrosion in FLiBe is provided in a report by Keiser [39]. This is consistent with our experiments, as the emissivity of 316 stainless steel experiences a significant increase after exposure to FLiBe (Fig. 5). To account for localized surface heterogeneities, four measurements were taken, and an average is presented, with the shaded region indicating the uncertainty. The roughness of the FLiBe samples, provided in Fig. 5 caption, is small compared to the measurement wavelengths, so scattering is unlikely to significantly affect the measurement. We note that the measurements of the unexposed sample matches well the data from refs [40,41].

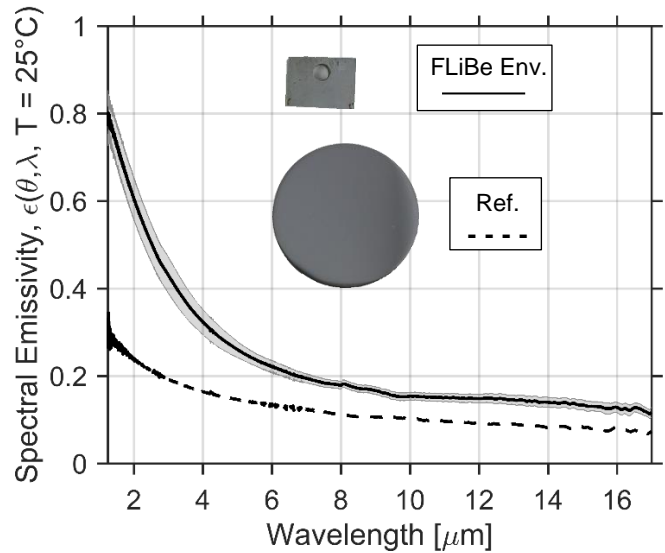

Fig. 5. Spectral normal emissivity for 316 stainless steel before and after exposure to 750 °C FLiBe for 1000 hours. Measurements were taken at room temperature using the reflective technique, with a numerical aperture of 0.4. Note that the wavelength range is slightly different from the other figures. The RMS roughness of the FLiBe-exposed sample was 0.20 µm, compared to 0.03 µm prior to exposure.

## 4.3 Liquid Sodium Environment

Stainless steel 316 is being considered for use in sodium-cooled fast reactors, which are expected to operate at temperatures as high as 700 °C [42]. Samples for this study were exposed to liquid sodium at 650 °C in a sodium loop facility for 1000 hours [22]. The oxygen concentration in the liquid sodium was 10 ppm, and the sodium flow velocity was 6 m/s.

Although decarburization of alloys after long-term exposures in molten sodium has been reported, the corrosion rate is generally considered to be quite low compared to other reactor-relevant environments [43]. Consistent with this, the emissivity of the exposed 316 stainless steel samples remained relatively unchanged after sodium exposure (Fig. 6).

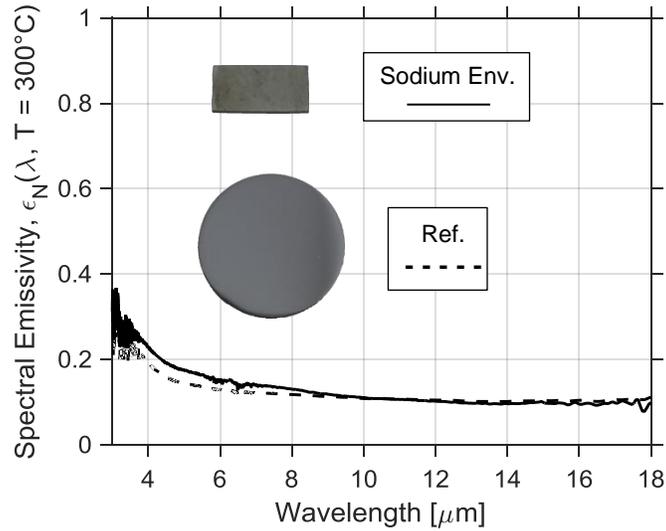

Fig. 6. Spectral normal emissivity for 316 stainless steel before and after 1000-hour exposure to liquid sodium at 650 °C. The measurements were taken at 300 °C. Photos of the samples are provided with the background cropped out. The exposed sample had an RMS roughness of 0.3 µm, compared to 0.03 µm prior to exposure.

## 4.4 Supercritical $CO_2$ Environment

The supercritical $CO_2$ (s$CO_2$) environment was selected because of the increasing interest in the s$CO_2$ Brayton cycle, which provides higher efficiencies than the steam Rankine cycle at 650 °C and above [44], and is therefore attractive for advanced reactor concepts. For this study, the samples were exposed to 650 °C s$CO_2$ at a pressure of 20 MPa and exposed to supercritical s$CO_2$ in a specially designed autoclave system [45]. A system-average flow rate of 0.11 kg/hr was maintained, resulting in a $CO_2$ refresh rate of one per two hours. $H_2O$, $N_2$, and $O_2$ impurities were monitored and regulated below a few ppm.

Fig. 7 shows the spectral emissivity measurements for Ni-based alloy Haynes 230 after exposure to s$CO_2$ at 650 °C. Based on previous exposure studies, the surface oxide layer is expected to be predominantly $Cr_2O_3$, with small amounts of $Cr_2MnO_4$ [23]. Our EDS measurements, provided in section 3.3 of the supplementary [27], indicate the presence of a chromium oxide with possible traces of manganese. Upon exposure, the spectral emissivity increases, but does not reach as high a value as the oxidizing-He-exposed Haynes 230 (Fig. 2). Similar to the oxidizing helium exposure case, a peak in the emissivity is observed for the Ni-based alloy between $16 < \lambda < 18$ µm. This is possibly associated with the absorption bands of $Cr_2O_3$, which occur near that region [32,46]. Visually, the exposed sample developed a spotty, greenish, partially reflective surface.

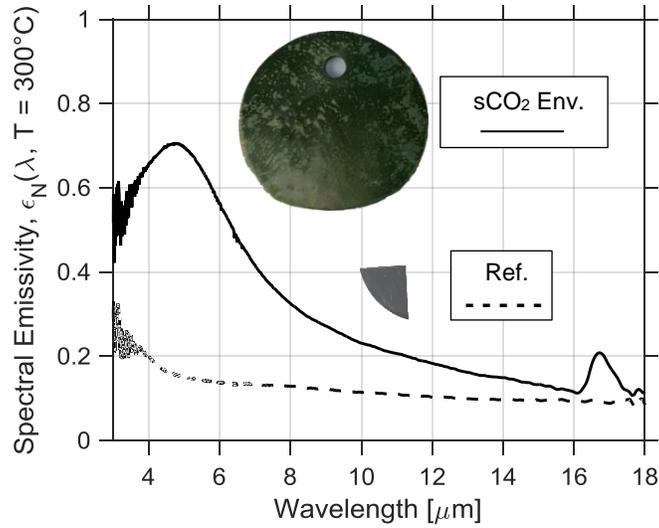

Fig. 7. Spectral normal emissivity for Haynes 230 after 400-hour exposure to 650 °C supercritical $CO_2$. Measurements taken at 300 °C. Photos of the samples are provided with the background cropped out. The exposed alloy had an RMS roughness of 0.4 µm, with 0.03 µm for the unexposed reference.

As shown in Fig. 8, 316 stainless steel experiences a much greater increase in emissivity than Haynes 230 under the same test conditions. As expected from previous literature [47], iron-based 316 stainless steel is more susceptible to oxidation than the nickel-based Haynes 230. Iron-rich oxide layers dominated the topmost surface layer in a previous high-temperature $sCO_2$ exposure study of 316 stainless steel under similar conditions to those in this study [48]. Another study reported multilayer oxide formation with iron-rich oxides on the outermost layer under related conditions [49].

Our surface grew a rough, grey surface layer which had a similar appearance to the surfaces of oxidized low-alloy steels [15]. EDS scans provided in section 3.4 of the supplementary [27] suggest an iron-oxide surface.

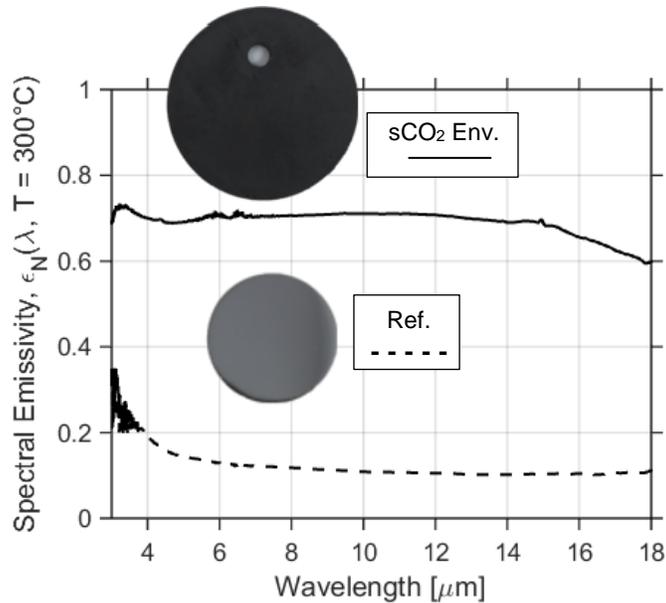

Fig. 8. Spectral normal emissivity for stainless steel 316 before and after 400-hour exposure to 650 °C supercritical $CO_2$. Measurements taken at 300 °C. Photos of the samples are provided on the figure with the background cropped out. The exposed sample had an RMS roughness of 1.6 µm, with 0.03 µm for the unexposed reference.

No obvious features are observed in the spectra of the sCO$_2$-exposed 316 stainless steel up to $\lambda = 18$ µm. If we assume iron-rich oxides form on the surface of this sample, the topmost oxides should either be magnetite ($Fe_3O_4$) [49] or a very thin layer of hematite ($Fe_2O_3$) on top of magnetite ($Fe_3O_4$) [48]. $Fe_2O_3$ exhibits two absorption peaks similar to those of $Cr_2O_3$ mentioned previously, but shifted to longer wavelengths by 2-5 µm [32], past our upper measurement limit of 18 µm. The resonant features of $Fe_3O_4$ occur around $16 < \lambda < 25$ µm [50–53]. We provide extended wavelength results in section 2 of our supplementary [27] and a peak is observed in this region; however, the uncertainty in these extended measurements is not quantified.

## 5 Conclusion

The emissivity of several reactor structural alloys before and after exposure to various advanced reactor coolant environments has been measured. We observed that the nature and extent of surface corrosion can have a significant effect on spectral emissivity. For Ni-based alloys Inconel 617 and Haynes 230, a thick black oxide layer developed on the surface after high-temperature exposure to helium gas with oxidative impurities, resulting in a large increase in emissivity. Similar effects were observed in high-temperature supercritical $CO_2$ environments, where oxidation-resistant alloy Haynes 230 had a smaller change in emissivity than 316 stainless steel. In contrast, an environment of helium gas with carburizing impurities resulted in only moderate increases in emissivity. For molten-fluoride-salt exposures where corrosion occurs by the dissolution of Cr from the alloy into the salt [20,36] and the oxide layer is unstable, the 316 stainless steel exhibited an increase in emissivity, while Hastelloy N exhibited almost no increase. Finally, exposure of 316 stainless steel to liquid sodium resulted in negligible change in emissivity because of the low corrosivity of this environment.

Our work provides a starting point for future investigations of alloy emissivity, and may be incorporated into models that better account for radiative heat transfer. Future works may more-precisely determine the composition, extent, roughness, and optical properties of the layers that are formed on the surface under various high-temperature conditions, e.g., using a combination of spectroscopic ellipsometry and cross-sectional electron microscopy, eventually enabling fully predictive models of the evolution of the emissivity of alloy surfaces in different reactor environments.

# 6 Acknowledgements

This research is supported by the U.S. Department of Energy Nuclear Energy University Program (NEUP) Contract No. DE-NE0000743, and by startup funds from UW-Madison.

We would like to acknowledge Robert Tirawat at the Concentrated Solar Power Group of the National Renewable Energy Laboratory for prior collaboration which enabled a useful comparison for our reflectance methodology. We also acknowledge Yuzhe Xiao for providing insight into the background contributions in the direct emission measurement setup, and Richard Wright for providing the helium-exposed nickel-based alloys measured in this study.

**Supplementary information:**

**Impact of Corrosion on the Emissivity of Advanced Reactor Structural Alloys**

J. L. King[1,2,3], H. Jo[2*], A. Shahsafi[1*]

K. Blomstrand[2], K. Sridharan[2]

M. A. Kats[1†]

[1]University of Wisconsin-Madison, Department of Electrical and Computer Engineering

[2]University of Wisconsin-Madison, Department of Engineering Physics

[3]Jensen Hughes, Power Services Group

[*]*These authors contributed equally to this work.* [†]*Corresponding Author: mkats@wisc.edu*

# 1  Uncertainty and Sources of Error

## 1.1  Direct Emission Measurement

Initially, it is important to verify the validity of the direct emission measurement method used in this study. Once the method is validated, uncertainty must be assigned to the measured emissivity values.

### 1.1.1  Validation of the Measurement Approach

We checked for various potential sources of measurement error. First, oxidation in air may occur during measurement, since the emission measurements are generally performed at elevated temperatures. Were this to happen, the room-temperature emissivity of a surface after an elevated-temperature measurement would differ from an earlier room-temperature measurement of the same surface. Examples of this hysteresis from oxidation can be found in high-temperature measurements of stainless steel taken by Rolling et al [S1]. Second, absorption and emission from ambient $CO_2$ and $H_2O$ (water vapor) may affect the measured signal. Lastly, background emission from the instrument, device surroundings, or some other unknown source can contribute to the measured signal. Each of these sources of uncertainty is discussed.

Oxidation during measurement was tracked by comparing the emissivity of samples before and after the elevated temperature measurement. Samples were always measured at 100 °C before and after each 300 °C measurement, and the two measurements at 100 °C were compared. No sample produced significantly different spectra between the two measurements.

To understand the contributions from the atmosphere, we measured the sample and the carbon nanotube (CNT) reference within at most a few hours of each other under the same controlled conditions. The reasoning behind this approach was that contribution from the atmosphere would approximately cancel out in Eq. S1 (Equation 7 in the main text) as long as the atmosphere was similar between the measurements of the sample and the blackbody reference.

$$\varepsilon(\lambda, T) = \varepsilon_R(\lambda, T) \frac{S(\lambda, T)}{S_R(\lambda, T)} \qquad (S1)$$

This approach required multiple measurements of the blackbody over the course of the study, and the measured signal of the blackbody reference could vary considerably within the wavelength regions of atmospheric absorption between different days. However, the resulting emissivity extracted using Eq. S1 remained consistent and reproducible with repeated measurements of the sample/reference pair.

Although the features from atmospheric absorption can still be observed in some measurements, they were greatly mitigated by taking measurements of the sample and the blackbody reference under very similar atmospheric conditions. In fact, the features associated with $H_2O$ are only visible in the emissivity spectra of nickel-based oxidative He-exposed alloys (seen in curve (i) in Fig. 2(a) and curve (i) in Fig. 3(a)). However, features due to $CO_2$ absorption could still be observed in the spectra of most samples. Fortunately, this region is isolated to a narrow range of 4.18 µm < λ < 4.4 µm. We assumed that the emissivity spectra of our samples did not have sharp features in this range and thus we excluded the data points, connected the remaining data with a straight line from the data point at λ = 4.18 µm to the point at λ = 4.4 µm ("linear spline").

The direct-emission measurement method used in the main text (referred to as the "simple method" for the remainder of this supplementary section) neglects any background emission from the instrument or from the surroundings. To evaluate the validity of this approach, we used two approaches that did incorporate background subtraction, and compared them to the simple method.

The first background subtraction method assumes that the background signal is temperature-dependent and sample-independent. This methodology, referred to here as the "gold subtraction method", assumed that the Fourier-transformed signal recorded by the FTIR followed Eq. S2:

$$S = \eta(\lambda)[\varepsilon(\lambda, T)I_{BB}(\lambda, T) + S_{BG}(\lambda, T)] \qquad (S2),$$

where $\eta(\lambda)$ is wavelength-dependent response function of the instrument and $S_{BG}$ is the background. For this method, $\eta$ was determined by measuring the blackbody reference sample at 300 °C and assuming the signal was almost entirely composed of the sample emission component of Eq. S2. If $S_R$ is measured and $\varepsilon_R(\lambda,T)I_{BB}(\lambda,T) \gg S_{BG}(\lambda,T)$ where $\varepsilon_R(\lambda,T)$ and $I_{BB}(\lambda,T)$ are known values, $\eta$ may be closely approximated by dividing $S_R$ by $\varepsilon_R(\lambda,T)I_{BB}(\lambda,T)$. Here, we made this assumption because our CNT reference was the most emissive sample we measured. If the assumption turned out to be invalid, the extracted emissivity using Eqs. 1 and 2 of samples with lower emissivity would differ substantially from each other. We found that this does not occur (see below).

Given Eq. S2, three objects are measured:

1) The sample of interest – $\varepsilon$ is denoted without subscript
2) A reflective gold film sample with $\varepsilon_{Au}(\lambda,T)$ close to 0. In our case, a 180-nm-thick gold coating was vapor-deposited onto a polished fused silica wafer
3) An additional emissive reference sample with a known emissivity, $\varepsilon_R(\lambda,T)$ (the same CNT reference used in the simple method)

Values of the gold reference sample are assumed to be similar to the literature values reported by Aksyutov [S2]. Each $\varepsilon_{Aksytov}(\lambda,T)$ is a fitting of Aksytov's data via linear interpolation, which was measured at $T = 300$ °C, the same temperature as the measurements performed in this study.

With the two reference samples, $\varepsilon(\lambda,T)$ is calculated according to Eq. S3:

$$\varepsilon(\lambda,T) \approx \varepsilon_R(\lambda,T) \frac{S-S_{Au}+\eta\varepsilon_{Aksyutov}(\lambda,T)I_{BB}(\lambda,T)}{S_R-S_{Au}+\eta\varepsilon_{Aksyutov}(\lambda,T)I_{BB}(\lambda,T)} \quad (S3)$$

Where $S$, $S_{Au}$, and $S_R$ are the measured emission from the sample of interest, the gold film, and the CNT reference, respectively.

The connection between the right and left sides of Eq. S3 can be shown by plugging in Eq. S2 into each $S$ on the right side of Eq. S3. If we assume $\varepsilon_{Au}(\lambda,T) \approx \varepsilon_{Akysutov}(\lambda,T)$ and that all temperatures and measured surface areas are equal, the right hand side of Eq. S3 collapses into $\varepsilon(\lambda,T)$, eliminating all background terms.

In Fig. S1, emissivities of two samples are calculated using both the simple method and the gold subtraction method. The samples selected for presentation here exhibited more-substantial differences in emissivity between the two methods compared to other samples we measured. Nonetheless, for these samples and for all other samples, the difference in emissivity between the two methods is mostly negligible.

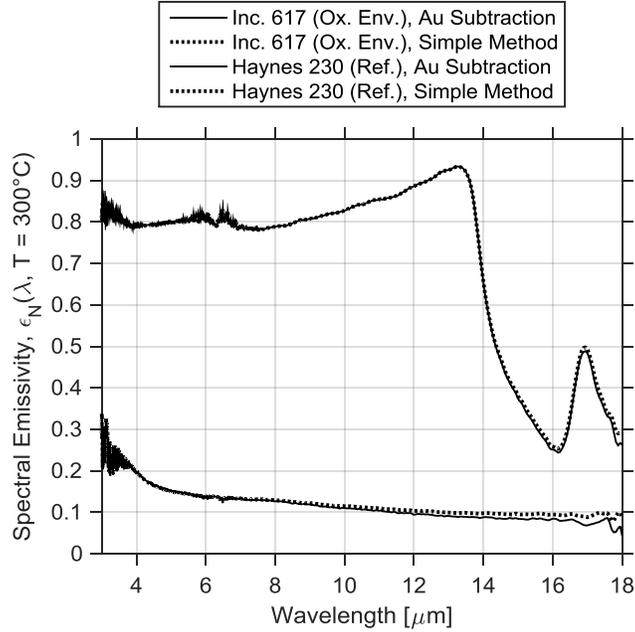

Fig. S1. Spectral normal emissivities for Inconel 617 after 500-hour exposure to 1000 °C helium gas with oxidative impurities, and reference Haynes 230, determined using the simple method and the gold subtraction method. In the most extreme case, emissivities differ by .04 between the two methods, but generally the difference is .01 or less. Simple method data for the Inconel 617 sample is taken from Fig. 3 of the main text while data from the Haynes 230 sample is taken from Fig. 2 of the main text.

The second background-subtracting method, referred to as the self-subtraction method, assumed the background to be sample-dependent (perhaps through the sample's capability to reflect or scatter light emitted elsewhere) and temperature-independent, following Eq. S4:

$$S = \eta(\lambda)[\varepsilon(\lambda,T)I_{BB}(\lambda,T) + \alpha(\lambda)\varepsilon(\lambda,T) + \beta(\lambda)] \qquad (S4)$$

Where $\alpha(\lambda)$ and $\beta(\lambda)$ are some unknown temperature-independent environmental background functions. This time, $\varepsilon(\lambda,T)$ is calculated according to Eq. S5:

$$\varepsilon(\lambda,T) \approx \varepsilon_R(\lambda,T)\frac{S(\lambda,T_1)-S(\lambda,T_2)}{S_R(\lambda,T_1)-S_R(\lambda,T_2)}, \qquad (S5)$$

where, for this study, $T_1 = 300$ °C and $T_2 = 100$ °C. Four measurements are taken, per one set of emissivity data. The reference sample is measured once at each of the two temperatures, and the sample of interest is measured once at each of the two temperatures. The assumption in this method is that spectral emissivity stays constant between the measurements taken at 100 °C and 300 °C. In Fig. S2, emissivities of two samples are calculated using both the simple method and the self-subtraction method. The samples selected for presentation here exhibit greater difference in emissivities calculated using the two methods compared to other samples. Differences are negligible up until the longest wavelengths, where signal-to-noise issues limit the performance of the self-subtraction method.

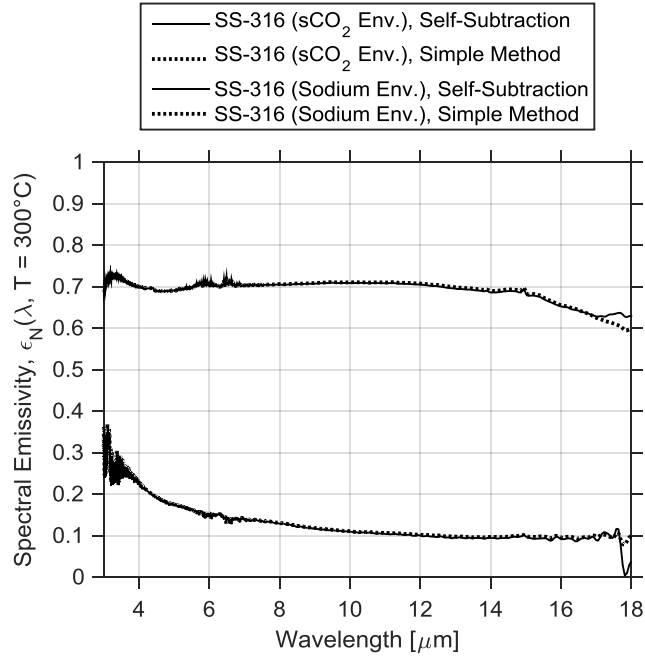

Fig. S2. Spectral normal emissivities for 316 stainless steel after 400-hour exposure to 650 °C supercritical $CO_2$ and 316 stainless steel after 1000-hour exposure to liquid sodium at 650 °C. Emissivities determined using the simple method and the self-subtraction method. Outcome is practically identical (< 1% relative difference) up until around 17.5 μm where difference may be attributed to poor signal to noise in the self-subtraction method. Simple method data for the sCO$_2$-exposed sample is taken from Fig. 8 of the main text while data from the sodium-exposed sample is taken from Fig. 6 of the main text.

### 1.1.2 Measurement Uncertainty

The simple method (Eq. S1 in supplementary and Eqs. 6-7 in the main text) can be expanded as:

$$\varepsilon(\lambda, T) = \varepsilon_R(\lambda, T) \frac{\eta(\lambda)\varepsilon(\lambda,T)I_{BB}(\lambda,T)}{\eta(\lambda)\varepsilon_R(\lambda,T)I_{BB}(\lambda,T)} \qquad (S6)$$

We consider the effects of uncertainty in $T$, $\varepsilon_R$, and $\eta(\lambda)$.

Note that the temperature may impact the spectral emissivity as well as the Planck distribution associated with a given emitting object. However, the potential deviations in temperature (between the blackbody reference sample and the sample of interest) in this study are on the order of a few degrees. While this change in a few degrees will noticeably impact the Planck distribution, it should have no measureable impact on the spectral emissivity, and so we consider it to be temperature-independent.

### 1.1.2.1  Variation in Surface Temperature

The heated stage temperature is fixed in the experimental setup, so samples of differing thermal resistance will also have differing surface temperature under the same conditions. A thick, insulating sample will have a slightly lower surface temperature than a thin, conducting sample for the same heated stage temperature as illustrated in Fig. S3 (exaggerated). This is especially true if the contacting surface area of Sample 1 is smaller than the contacting surface area of Sample 2.

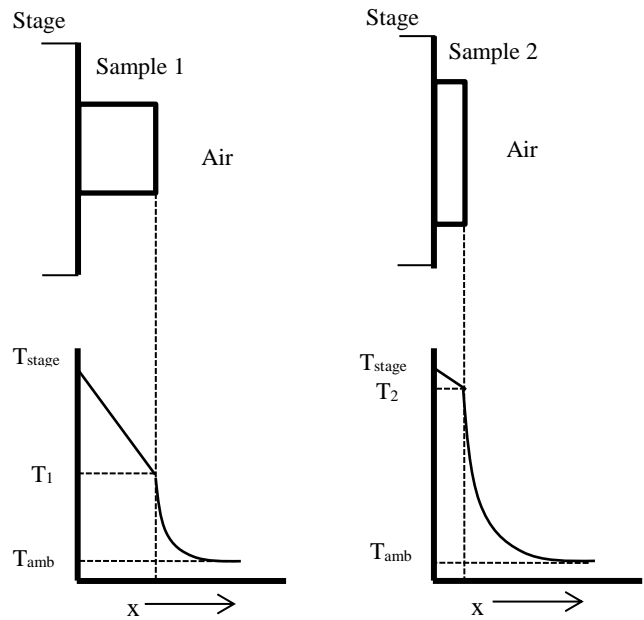

Fig. S3. Exaggerated temperature profiles of a thick insulating sample (left) and a thin conducting sample (right) attached to the device stage. The measured radiative power is a function of surface temperature ($T_1$ and $T_2$). $T_{amb}$ represents the ambient temperature.

The potential difference in surface temperature is approximated from heat transfer calculations. The system can be described by sample thermal conductivity ($k$), stage temperature ($T_{stage}$), sample thickness ($x$), and some heat transfer coefficient from the sample surface to the surrounding air ($h$). We assume the thermal resistance of the thermal paste to be negligible.

The system was treated as a vertical surface experiencing free convection with room-temperature air. The associated Prandtl ($Pr = 7.03\text{E-}1$), Rayleigh ($2.01\text{E}2 < Ra < 4.34\text{E}4$, varying by sample), and Nusselt numbers ($2.6 < Nu < 8.1$, varying by sample) of the system were used along with equations developed by Churchill and Chu [S3] to determine $h$ that describes convection ($11.8 < h < 22.9$ W/m$^2$-K). This value was added to the radiative heat transfer coefficient. The overall heat transfer coefficient, including convection and radiation, is set to a value of 50 W/m$^2$-K, slightly larger than the sum of the two calculated *h* values (one for convection and one for radiation) under maximal conditions.

The error in the measured emissivity can arise from surface-temperature differences between the sample and the CNT reference. Based on material parameters and dimensions of all samples measured, the CNT reference is expected to have the lowest thermal resistance (it was the thinnest sample and was composed of thermally conductive silicon and carbon nanotubes), while the oxidizing-He-exposed Inconel 617 is expected to have the highest thermal resistance (it was the thickest sample and also had an thermally insulating oxide layer). Using basic heat transfer methods, we compared the expected surface temperature of the reference to that of a hypothetical Inconel 617 sample having greater thermal resistance than that expected from our actual sample. The hypothetical Inconel 617 had a thickness of 3.9 mm (actual sample had a thickness of 3 mm) and a $Cr_2O_3$ oxide layer thickness of 50 μm (the actual sample should have an oxide layer of around 20 - 30 μm based on analysis performed by Wright [S4]). A surface temperature difference of 4 °C was calculated between the CNT reference and the hypothetical Inconel 617, so we expect $\Delta T = 4$ °C to be the upper bound for the temperature difference between the surface of any sample and the surface of the CNT reference.

According to Eq. S6, differences in surface temperature between the sample and the CNT reference result in uncertainty in calculated emissivity due to differences in the temperature-dependent Planck distributions. This difference is carried through by using the term *F* and *ΔF*, which we define as

$$F = \frac{I_{BB}(\lambda,T)}{I_{BB}(\lambda,T_R)} \quad (S7)$$

and

$$\Delta F = |1 - F| \quad (S8),$$

where *T* is the surface temperature of the sample and $T_R$ is the surface temperature of the CNT reference. For a given $\Delta T$ (temperature difference the sample and the CNT reference) and approximately the same central temperature, *F* is relatively unchanged. For example, for $\Delta T = 4$ °C,

$$\frac{I_{BB}(\lambda, T = 285\ °C)}{I_{BB}(\lambda, T = 289\ °C)} \cong \frac{I_{BB}(\lambda, T = 295\ °C)}{I_{BB}(\lambda, T = 299\ °C)}$$

For this reason, *F* and $\Delta F$ can be viewed as functions of *λ*, *T* and *ΔT*, where, for this study, *T* is $T_{Stage}$ and $\Delta T = 4$ °C. Curves of $\Delta F$ may be easily generated using the Planck distribution for several different conditions. Examples of $\Delta F(\lambda, T = 300\ °C, \Delta T)$ are given in Fig. S4. The ΔT= 4 °C curve represents the present study.

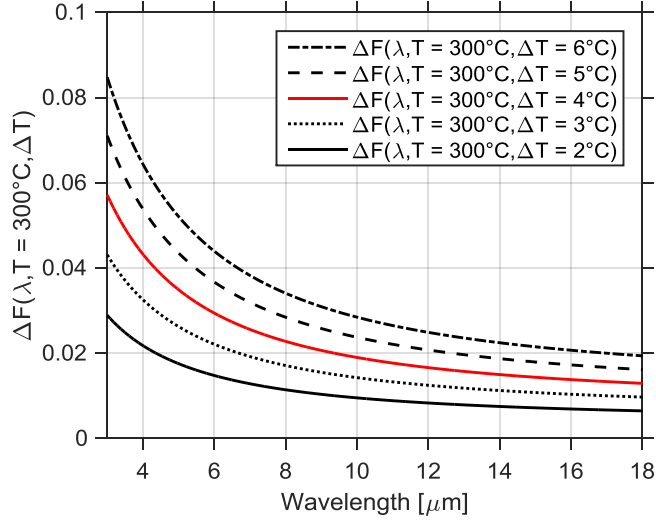

Fig. S4. $\Delta F(\lambda, T = 300\ °C, \Delta T)$. The largest errors related to temperature differences are expected at shorter wavelengths.

### 1.1.2.2 Error in Reference Emissivity

Vertically-aligned carbon nanotube forests of sufficient height have proven to be capable of absorbing light across a wide spectral range (0.2-200 μm) [S5], and so they make effective laboratory blackbody references. Our reference was a vertically aligned carbon nanotube array grown by NanoTechLabs Inc. Carbon nanotubes were grown on a silicon wafer to a height of 100 μm. This thickness turns out to not be sufficient to achieve ideal blackbody absorption/emissivity, but our calibrations (described below) showed that this reference had a uniform spectral emissivity of 0.90 +/- .02 throughout the measured wavelength range.

The reference sample was calibrated by comparing measurements using the direct emission measurement technique to measurements using the reflectance technique (and Kirchhoff's law) of two well-known polished surfaces (sapphire and fused silica). First, the reflectance technique was used to acquire an emissivity value (via Eq. 8 of the main text) for the sapphire sample. Then, based on Eq. 7 of the main text, the emission of the sapphire and reference were measured, and the input value for $\varepsilon_R$ was adjusted until good agreement of the resultant sample $\varepsilon$ was achieved between the two techniques. The emissivities acquired from these two methods for the sapphire sample are provided in Fig. S5a. Using this new $\varepsilon_R$, measurements of the emissivity of fused silica were compared between the two methods, this time without adjusting $\varepsilon_R$. The emissivities acquired from the two methods for the silica sample are provided in Figure S5b, and are very consistent. The $\varepsilon_R$ determined in this manner was .90 +/- .02 across all measured wavelengths.

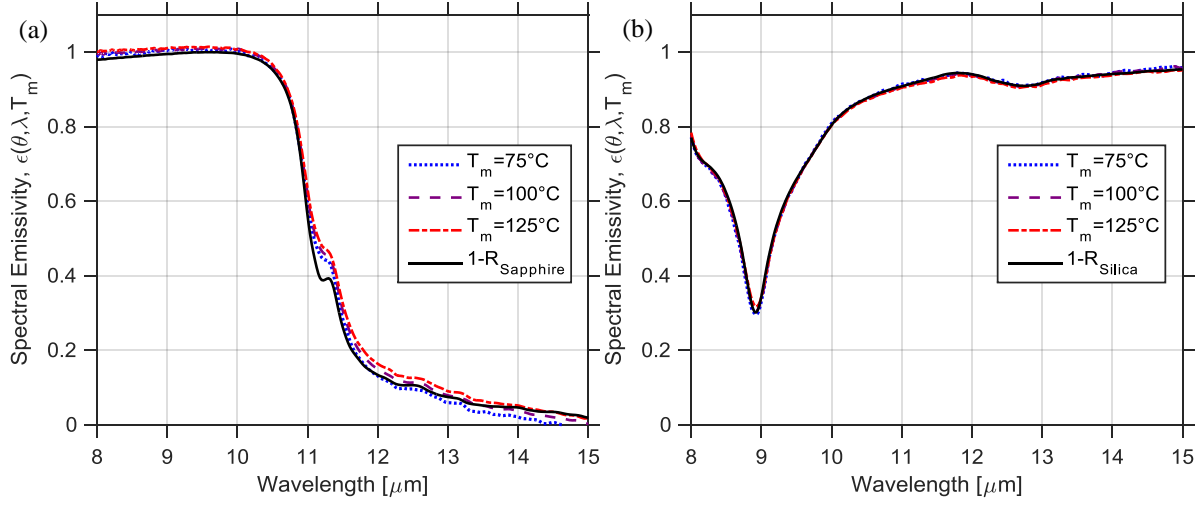

Fig. S5. Spectral emissivity of sapphire (a) and silica (b) measured using direct emission (dotted, dashed, and dot-dashed) and reflection (solid). Calibration of the emissive reference is performed based on this measurement, resulting in $\varepsilon_R = 0.90$ +/- .02. $T_m$ is measurement temperature.

In every case, the sapphire and silica samples were sufficiently thick and flat to assume negligible transmission and scattering. Measurements were taken at several temperatures in the case of the direct emission measurements to verify consistency. Measurements using the reflectance technique were performed at room temperature. Note that the reflection technique could not be used to directly calibrate the reference, because the CNT forest may not be sufficiently flat to suppress all scattering.

### 1.1.2.3 Overall Uncertainty

Standard uncertainty evaluation was performed, assuming that the dominant sources of error are (1) surface temperature differences between the reference and the sample of interest, and (2) the uncertainty in the emissivity of the reference sample. The total uncertainty can be evaluated using error propagation, yielding:

$$\Delta\varepsilon_{measure}(\lambda, T, \Delta T) = \varepsilon_{measure}\sqrt{2\left(\frac{\Delta\varepsilon_R}{\varepsilon_R}\right)^2 + \Delta F(\lambda, T, \Delta T)^2} \quad (S9)$$

The square root term is defined as the relative overall error factor, $Q$

$$Q \equiv \sqrt{2\left(\frac{\Delta\varepsilon_R}{\varepsilon_R}\right)^2 + \Delta F(\lambda, T, \Delta T)^2} \quad (S10)$$

Using $\Delta\varepsilon_R/\varepsilon_R = .023$ and the equation for $\Delta F(\lambda, T = 300\ °C, \Delta T)$, $Q$ is evaluated and plotted in Fig. S6 for several different $\Delta T$. $Q(\lambda, T = 300\ °C, \Delta T = 4\ °C)$ corresponds to this study.

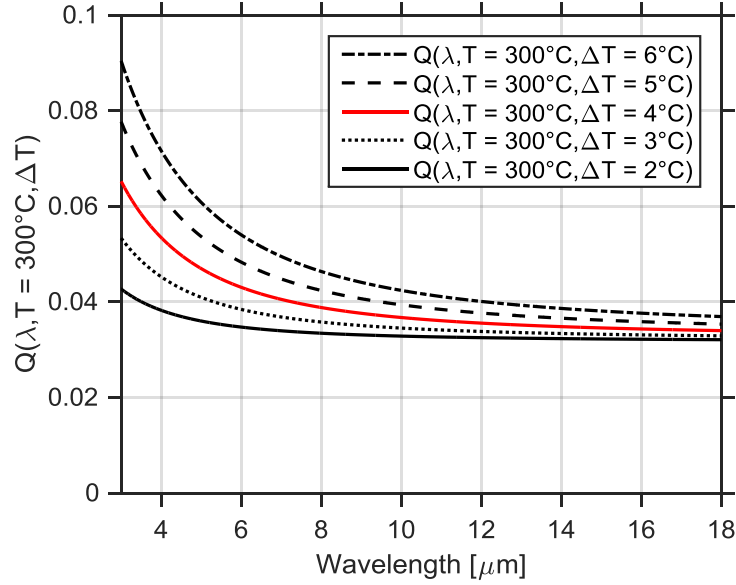

Fig. S6. Relative overall error profile for direct emission measurements, $Q(\lambda, T = 300\ °C, \Delta T)$. $F(\lambda, T, \Delta T)$ contributes more significantly to uncertainty at shorter wavelengths. $Q(\lambda, T = 300\ °C, \Delta T = 4\ °C)$ corresponds to this study and is plotted in red.

The confidence region of the emissivity is thus $\varepsilon(\lambda, \theta_N) = \varepsilon_m(\lambda, \theta_N) \pm Q(\lambda, \Delta T)\, \varepsilon_m(\lambda, \theta_N)$ for each sample, where $\varepsilon_m$ is the measured emissivity.

## 1.2 FTIR Reflectance Measurements

Unlike emission measurements, the reflectance measurement does not require elevated temperature to generate adequate signal. At low temperatures, in-situ oxidation is no longer a concern. However, the reflection method is valid for only non-transmitting media and may be invalid for rough or scattering media unless the roughness is substantially small, or all scattered light is collected. To make sure that scattering did not substantially affect our measurement for a FLiBe-exposed sample that was too small to measure any other way, we performed a test of the reflection method using a roughened steel sample that could be measured with both methods. Reflectance from the Bruker Vertex 70 FTIR was calibrated using a silicon dioxide wafer with a coating of vapor-deposited gold assuming a reflectance of R = .98 +/- .01.

As a part of another study [S6], a large, low-alloy steel reference sample with slightly greater RMS roughness (0.29 μm) and heterogeneity than the FLiBe-exposed sample (RMS roughness of 0.20 μm) had been measured using two other facilities that did not have roughness limitations. These facilities included a SOC-100 HDR FTIR system managed by the Concentrated Solar Power Group at the National Renewable Energy Laboratory (NREL), and a Surface Optics ET-100 Emissometer. The measurements using the three instruments are given in Fig. S7 for the low alloy steel reference. The observation of reasonable agreement for the low alloy steel sample validates the usage of Bruker Vertex FTIR with the reflectance technique for the FLiBe-exposed sample.

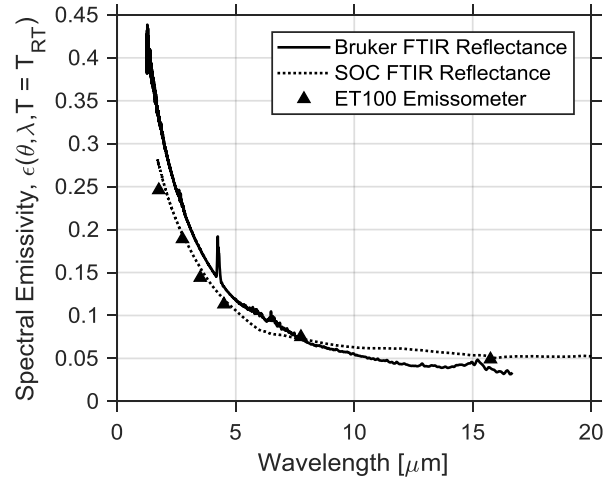

Fig. S7. Low-angle spectral emissivity of 120-grit-roughened SA508 low-alloy steel measured using three different experimental setups. The Bruker Vertex reflectance (solid line) was performed using a microscope (NA = 0.4). The SOC-HDR FTIR measurement (dotted line) involved directional-hemispherical reflectance measurement of diffusely and specularly reflected light at 10-degree incidence. The ET100 emissometer measurement (triangles) involve the measurement of diffuse and specularly reflected light using an integrating sphere from light at 20-degree incidence. All measurements were taken at room temperature.

## 2   Extended Spectra

Our instruments were able to capture a wider spectral range than what is presented in the main text, but the signal outside of $3 < \lambda < 18$ μm was poor, and we did not perform careful calibration of our measurements outside of this range.

Here we present an extended spectrum of one of our samples. Looking at the extended emissivity spectrum for the sCO$_2$-exposed stainless steel 316 (Fig. S8), we see a peak ($\lambda \sim 20$ μm) where we would expect resonant features to reside for iron oxides [S7,S8].

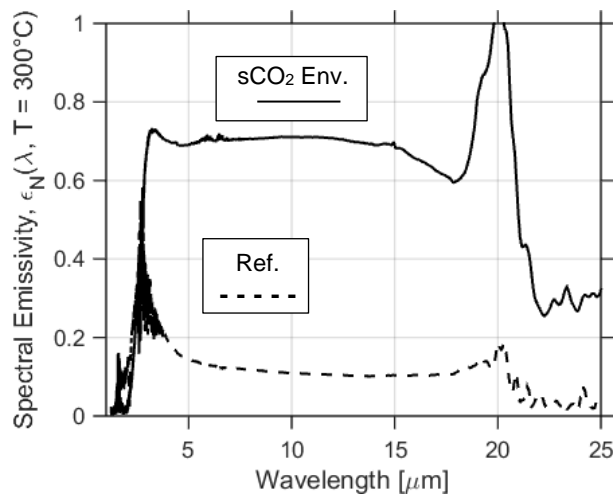

Fig. S8. Spectral normal emissivity for stainless steel 316 after 400-hour exposure to 650 °C supercritical $CO_2$. Measurements taken at 300 °C. An extended measurement range of $1 < \lambda < 25$ μm is provided.

Note that this data must be interpreted skeptically; *e.g.*, the emissivity cannot be greater than 1, as it appears to be around $\lambda = 20$ μm.

In our previous work [S9,S10], material resonances of varying magnitude appear in the spectral emissivity of oxidized low-alloy SA508 steel around the wavelength range of $17.5 < \lambda < 24$ μm. These SA508 alloys form iron-rich oxides on the surface and the material resonances observed in their spectra may be associated with absorption bands of iron oxides which occur around these wavelength regions. We postulate that the peak that occurs in the spectra of sCO₂-exposed stainless steel at $\lambda \sim 20$ μm is associated with resonances of iron oxide.

## 3 EDS Measurements

We used EDS to examine surface composition of the exposed samples. These samples included the He-exposed nickel-based alloys and the supercritical CO₂-exposed alloys.

### 3.1 He-exposed Haynes 230

After exposure to the carburizing helium environment, carbon and chromium peaks grew in relation to the nickel peaks (Fig. S9b). This suggests the formation of a chromium carbide on the sample surface. After exposure to the oxidizing helium environment, nickel peaks are no longer visible while chromium peaks remain pronounced (Fig. S9c). This suggests the formation of a chromium-dominant oxide.

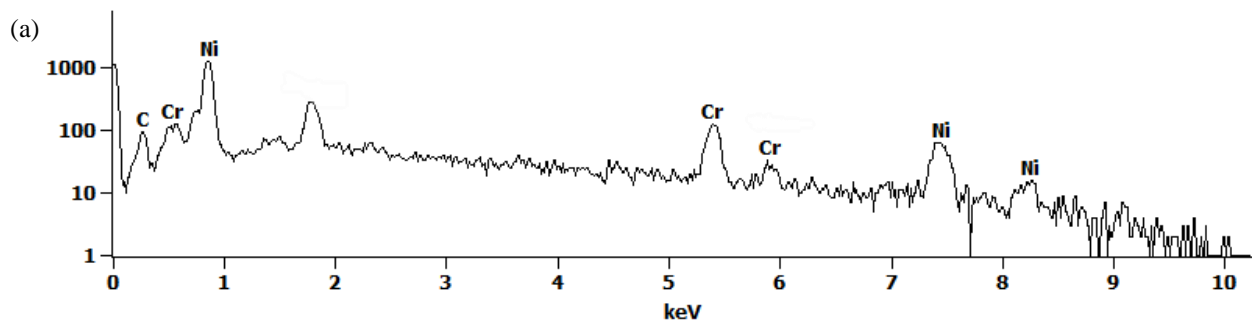
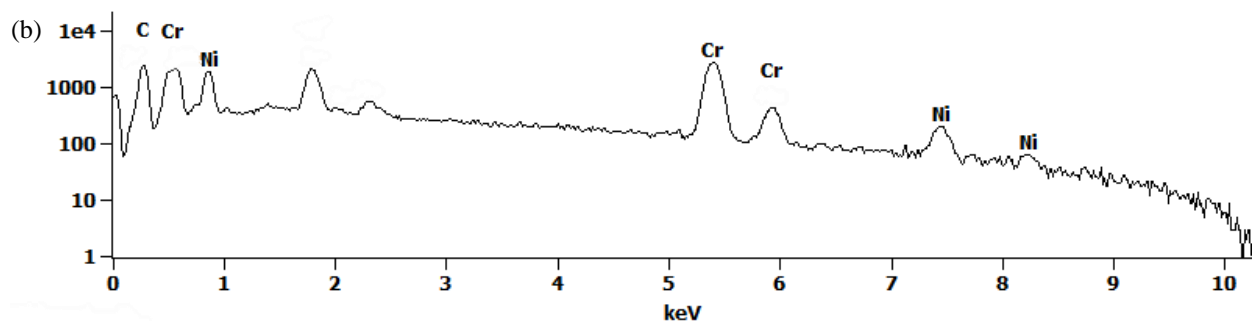
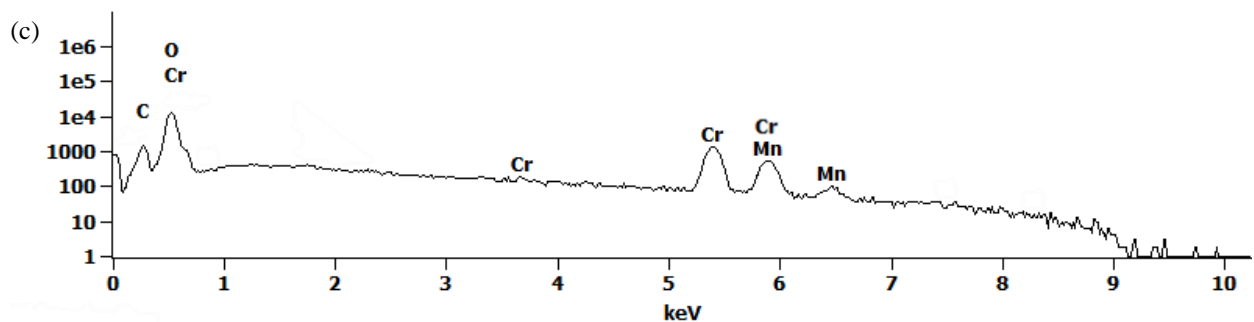

Fig. S9. EDS scans of (a) Haynes 230 reference sample, (b) Haynes 230 after 500-hour exposure to 1000 °C helium gas with carburizing chemistry, (c) Haynes 230 after 500-hour exposure to 1000 °C helium gas with oxidizing chemistry. These scans are for the samples presented in Fig. 2 of the manuscript.

## 3.2  He-exposed Inconel 617

EDS results of the Inconel 617 samples mirrored the EDS results of the Haynes 230 samples. After exposure to the carburizing helium environment, carbon and chromium peaks grew in relation to the nickel peaks (Fig. S10b). This suggests the formation of a chromium carbide on the sample surface. After exposure to the oxidizing helium environment, nickel peaks are nearly absent while chromium peaks remain pronounced (Fig. S10c). This suggests the formation of a chromium-dominant oxide.

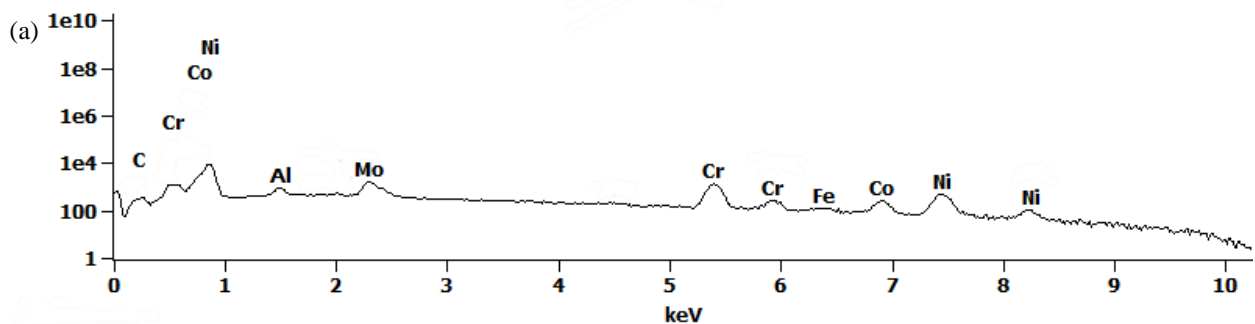

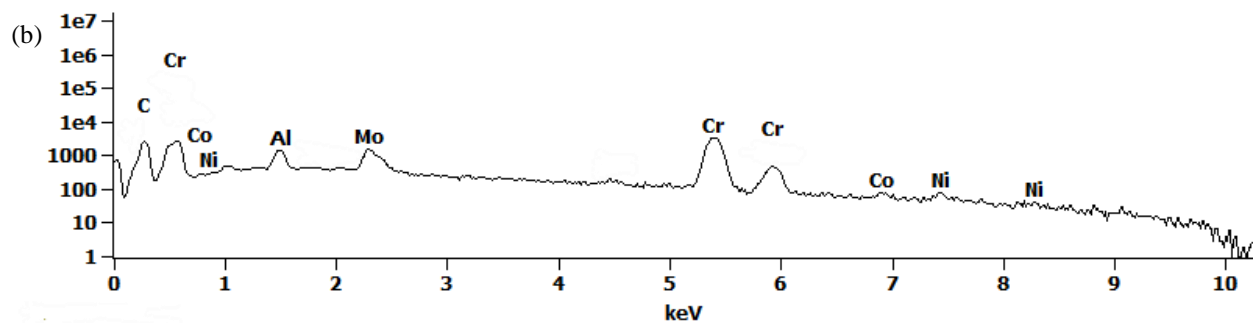

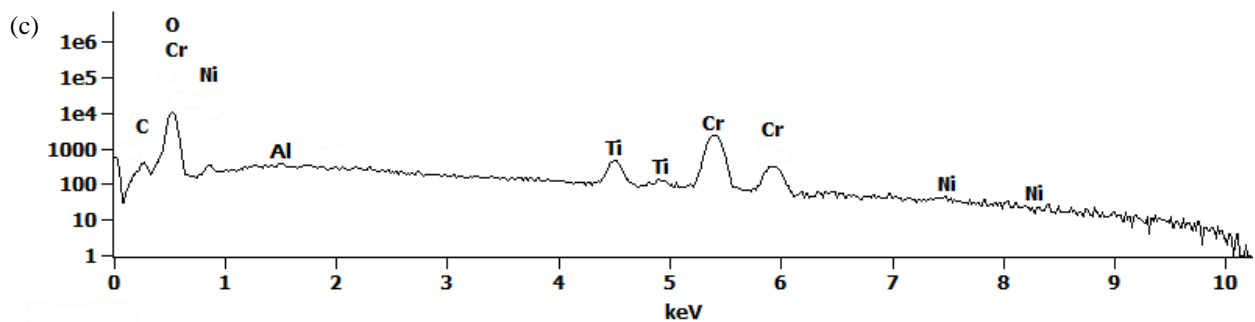

Fig. S10. EDS scans of (a) Inconel 617 reference sample, (b) Inconel 617 after 500-hour exposure to 1000 °C helium gas with carburizing chemistry, (c) Inconel 617 after 500-hour exposure to 1000 °C helium gas with oxidizing chemistry. These scans are for the samples presented in Fig. 3 of the manuscript.

## 3.3  FLiNaK-exposed Hastelloy N

After FLiNaK exposure, chromium and nickel peaks are subdued while peaks which seem to be associated with Zirconium appear (Fig. S11b). This is most likely due to plating of Zirconium impurities within the salt onto the surface of the Hastelloy N sample during exposure.

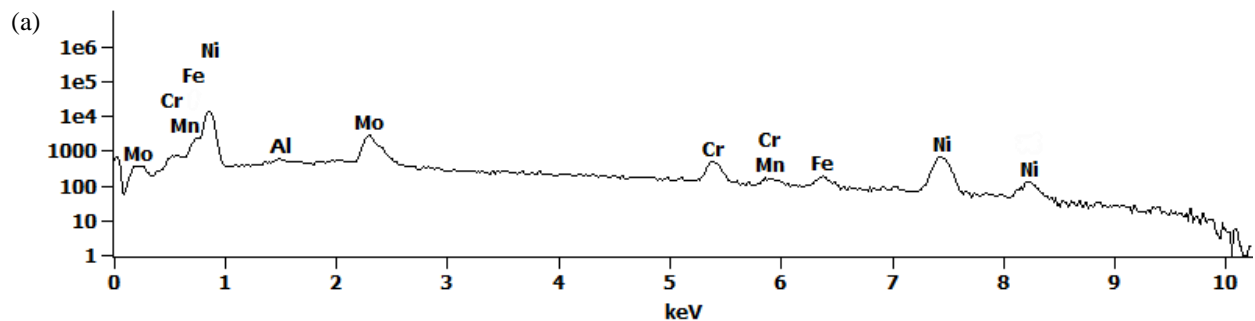

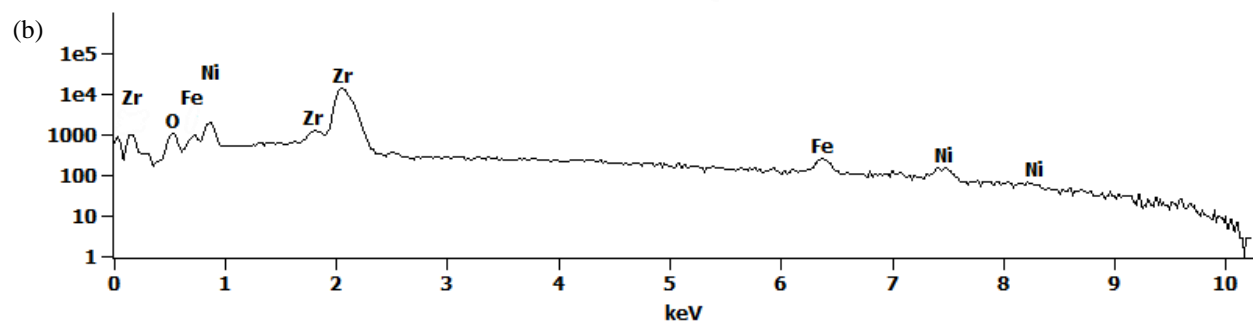

Fig. S11. EDS scans of (a) Hastelloy N reference sample, (b) Hastelloy N after exposure to 850 °C FLiNaK for 1000 hours. These scans are for the samples presented in Fig. 4 of the manuscript.

## 3.4 FLiBe-exposed 316 Stainless Steel

After FLiBe exposure, EDS spectra remains relatively unchanged. Chromium peaks appear to be slightly subdued (Fig. S12b).

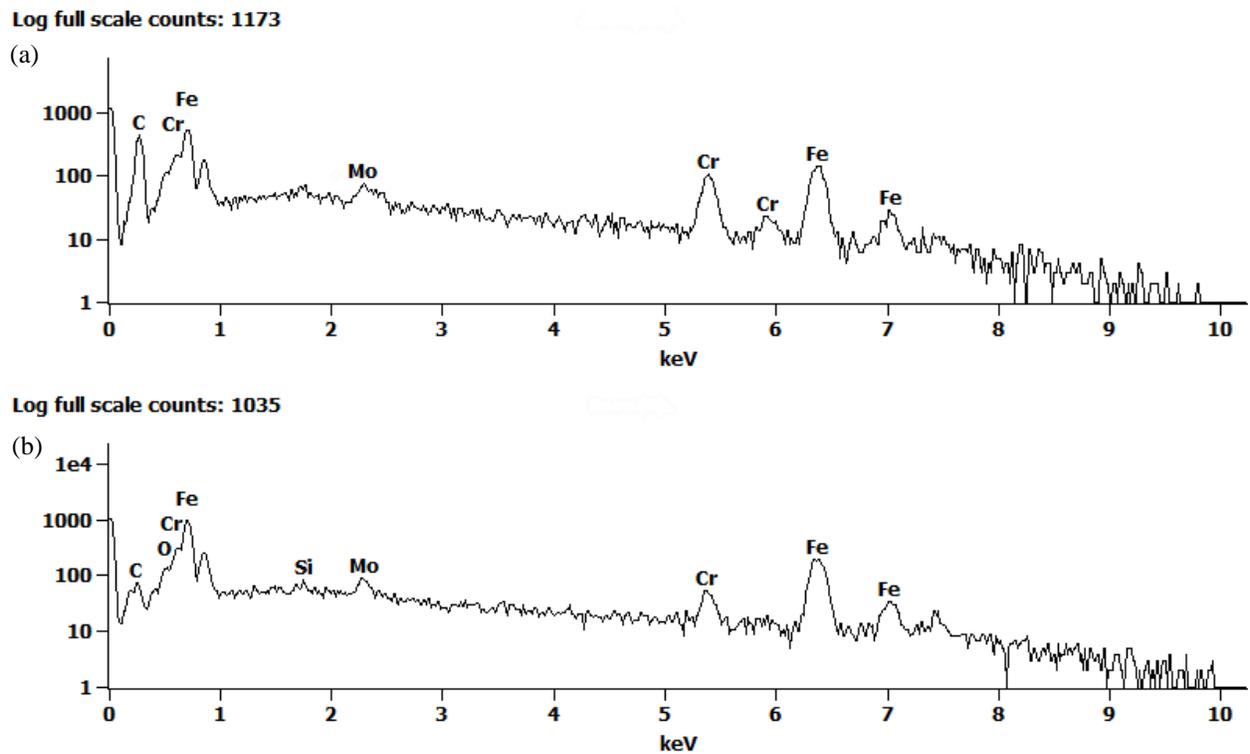

Fig. S12. EDS scans of (a) 316 stainless steel reference sample, (b) 316 stainless steel after exposure to 750 °C FLiBe for 1000 hours. These scans are for the samples presented in Fig. 5 of the manuscript.

## 3.5 Liquid Na-exposed 316 stainless steel

After Liquid sodium exposure, EDS spectra remains relatively unchanged (Fig. S13).

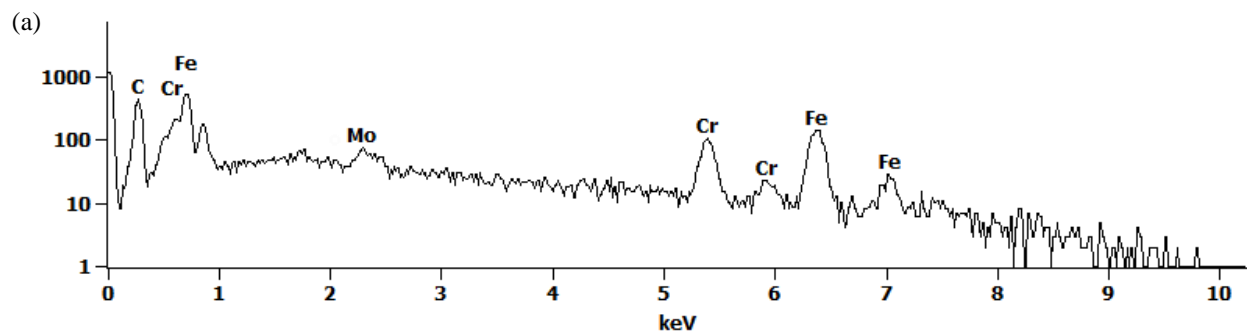

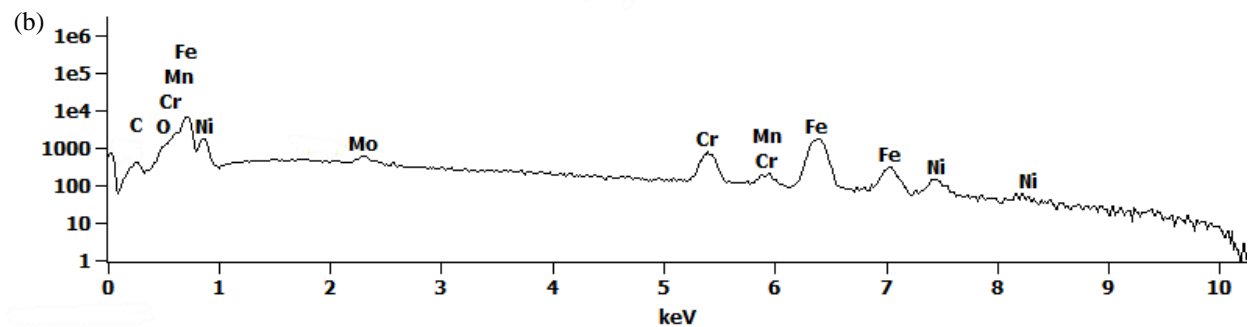

Fig. S13. EDS scans of (a) 316 stainless steel reference sample, (b) 316 stainless steel after 1000-hour exposure to liquid sodium at 650 °C. These scans are for the samples presented in Fig. 6 of the manuscript.

### 3.6  Supercritical $CO_2$-exposed Haynes 230

After $sCO_2$ exposure, nickel peaks are no longer visible while chromium peaks remain pronounced (Fig. S14b). This suggests the formation of a chromium-dominant oxide.

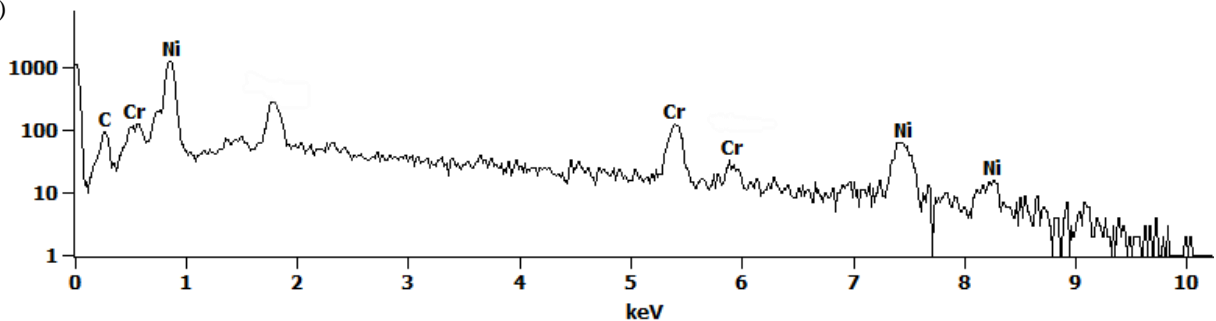

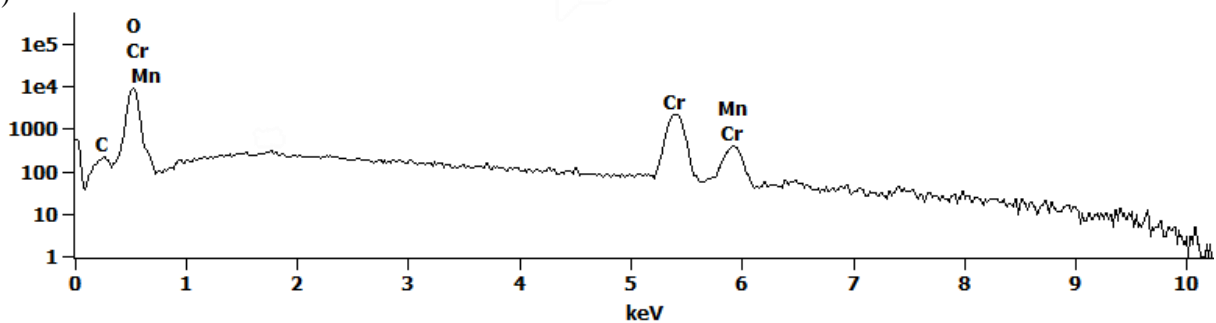

Fig S14. EDS scans of (a) Haynes 230 reference sample, (b) Haynes 230 after 400-hour exposure to 650 °C supercritical $CO_2$. These scans are for the samples presented in Fig. 7 of the manuscript.

## 3.7 Supercritical $CO_2$-exposed 316 Stainless Steel

After $sCO_2$ exposure, chromium peaks are no longer visible while iron peaks remain pronounced (Fig. S15b). This suggests the formation of an iron oxide on the surface.

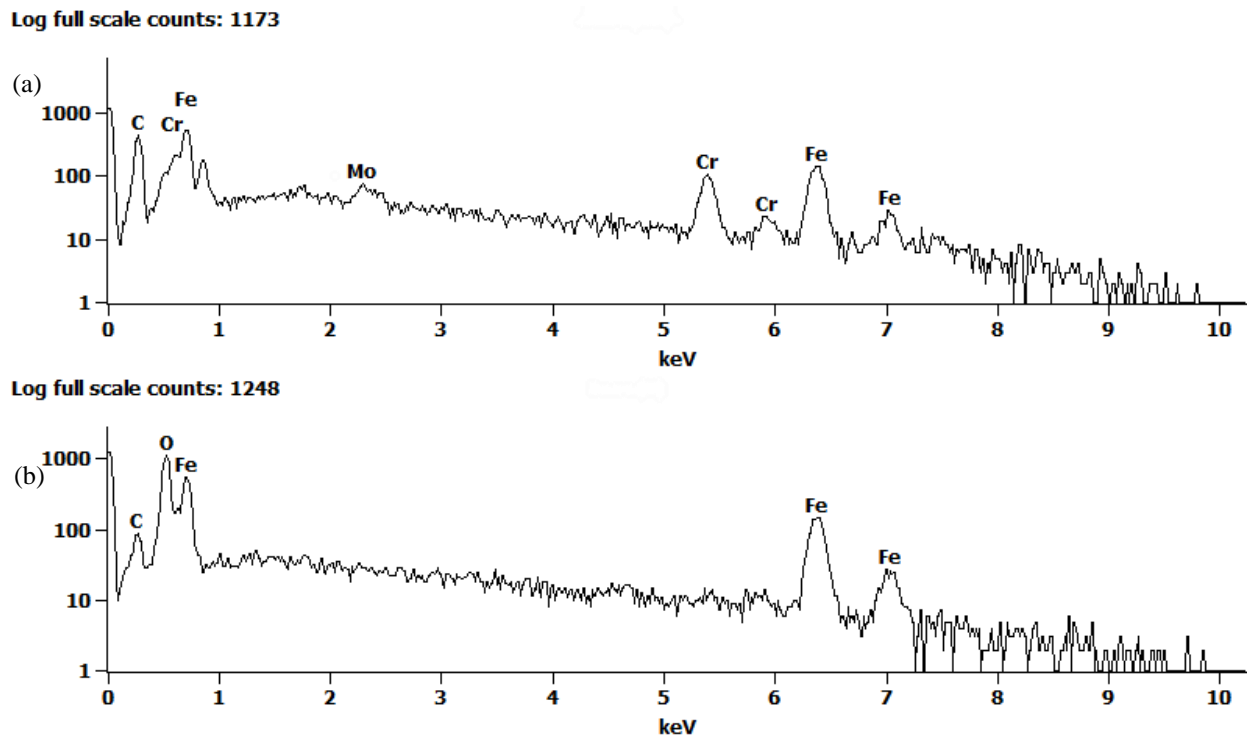

Fig S15. EDS scans of (a) 316 stainless steel reference sample, (b) 316 stainless steel after 400-hour exposure to 650 °C supercritical $CO_2$. These scans are for the samples presented in Fig. 8 of the manuscript.

# Supplementary references